\documentclass[twocolumn,secnumarabic,amssymb, nobibnotes, superscriptaddress, aps, prl]{revtex4-1}
\usepackage{hyperref}
\usepackage{amsmath}
\usepackage{graphicx}
\usepackage[dvipsnames]{xcolor}

\begin{document}

\title{All-optical blast wave control of laser wakefield acceleration in near critical plasma
}

\author{I. Tsymbalov}
\email{ivankrupenin2@gmail.com}
\affiliation{Faculty of Physics, Lomonosov Moscow State University, 119991, Moscow, Russia \looseness=-1}
\affiliation{Institute for Nuclear Research of Russian Academy of Sciences, 117312, Moscow, Russia \looseness=-1}

\author{D. Gorlova}
\affiliation{Faculty of Physics, Lomonosov Moscow State University, 119991, Moscow, Russia \looseness=-1}
\affiliation{Institute for Nuclear Research of Russian Academy of Sciences, 117312, Moscow, Russia \looseness=-1}

\author{K. Ivanov}
\affiliation{Faculty of Physics, Lomonosov Moscow State University, 119991, Moscow, Russia \looseness=-1}
\affiliation{Lebedev Physical Institute of Russian Academy of Sciences, 119991, Moscow, Russia \looseness=-1}

\author{E. Starodubtseva}
\affiliation{Faculty of Physics, Lomonosov Moscow State University, 119991, Moscow, Russia \looseness=-1}

\author{R. Volkov}
\affiliation{Faculty of Physics, Lomonosov Moscow State University, 119991, Moscow, Russia \looseness=-1}

\author{I. Tsygvintsev}
\affiliation{Keldysh Institute of Applied Mathematics of Russian Academy of Sciences, 125047 Moscow, Russia \looseness=-1}

\author{Yu.  Kochetkov}
\affiliation{National Research Nuclear University MEPhI, 115409, Moscow, Russia \looseness=-1}

\author{Ph. Korneev}
\affiliation{National Research Nuclear University MEPhI, 115409, Moscow, Russia \looseness=-1}

\author{A. Polonski}
\affiliation{Institute for Nuclear Research of Russian Academy of Sciences, 117312, Moscow, Russia \looseness=-1}

\author{A. Savel'ev}
\affiliation{Faculty of Physics, Lomonosov Moscow State University, 119991, Moscow, Russia \looseness=-1}
\affiliation{Lebedev Physical Institute of Russian Academy of Sciences, 119991, Moscow, Russia \looseness=-1}

\date{March 2024}%

\begin{abstract}

We propose a novel method for changing the length of laser wakefield electron acceleration in a gas jet by a cylindrical blast wave created by a perpendicularly focused nanosecond laser pulse. The shock front destroys the wake thus stopping interaction between the laser pulse and accelerated electron bunch allowing one to directly control the interaction length and avoid dephasing. It also improves the electron beam quality through the plasma lensing effect between the two shock fronts. We demonstrated both experimentally and numerically how this approach can be used to form quasi-monoenergetic electron bunch with controlled energy and improved divergence as well as to track changes in the bunch parameters during acceleration.

\end{abstract}
\maketitle

Laser-plasma electron accelerators currently achieve the generation of GeV-scale quasi-monochromatic low divergence electron beams using Laser Wakefield Acceleration (LWFA) in rarefied plasma and PW class lasers \cite{PhysRevLett.122.084801,10.1063/5.0161687}. At the same time, the generation of high repetition rate electron beams with energies of 5-50 MeV using denser, near critical plasma and shorter laser pulses with peak power of the order of 10 TW are also the focus of current studies \cite{salehi2017mev,Faure_2019,gustas2018high,lin2020laser}. Further, we will focus on such a case. 

The main stages of the LWFA are injection of electrons,  acceleration in the field of the plasma wake, and extraction of the beam from the plasma target. The extraction must be carried out before the onset of dephasing \cite{esarey2009physics}  to avoid a reduction in energy and an increase in divergence of the electron beam.
The onset of dephasing can be delayed or even canceled by controlling the evolution of the plasma wave using chirped \cite{Kalmykov_2012} or spatiotemporally shaped \cite{PhysRevLett.124.134802,PhysRevX.9.031044} laser pulses. Dephasing can also be avoided by terminating acceleration when electrons pass through the point where the wakefield equals zero or by matching laser pulse depletion length to the dephasing length\cite{Faure_2019}. 

Another method to avoid dephasing is to create a plasma profile that terminates at the desired stage of interaction.
To extract the electron beam, it is necessary to interrupt the acceleration process at least on a scale of $L_d/4$ ($L_d$ - dephasing length, i.e. length at which the phase of the plasma wave relative to the accelerated electron changes by $\pi$). For a rarefied plasma (electron density $n=0.01-0.001n_{cr}$) the dephasing length can be estimated  as $L_d \approx \lambda (n_{cr}/n)^{3/2}$ \cite{esarey2009physics} ($\lambda$ - laser wavelength, $n_{cr}$ - critical density) and amounts to more than $1000\lambda$. This is well matched with a density drop achieved at the edge of a supersonic gas jet \cite{liu2021effect} conventionally used for a GeV LWFA with low repetition rate PW class lasers. 

However, for high-repetition-rate systems with 10-100 times lower peak power, self-focusing is needed to produce elongated acceleration plasma channel  \cite{Faure_2019}, which requires higher plasma densities ($n\approx0.05n_{cr}$ for a 1 TW laser). For the density of $n\approx0.05n_{cr}$, the dephasing length can be estimated as $L_d \approx 90\lambda\approx 20\lambda_p$, $\lambda_p$ is plasma wavelength ($\lambda_p\approx 4.4\lambda$ for $n\approx0.05n_{cr}$). Thus, to suppress dephasing the plasma needs to be terminated at a scale of $5\lambda_p\approx22\lambda$.
Furthermore, the terminate the acceleration process on a scale of several $\lambda_p$ is very important as a diagnostic tool that allows direct investigation into the dynamics of electrons' energy gain. 

To terminate the interaction of the electron bunch with accelerating plasma wake density drop by an order on a scale of $10-20\lambda$ can be used.
Shock waves from an obstacle, such as a needle or blade inserted into a supersonic gas jet, are widely used to create a density ramp for injection of electrons. Typically, this results in a density drop by a factor of 1.5 -- 3  over $\approx100$ $\mu$m \cite{Fan-Chiang_2020,thaury2015shock}.
Density drop by an order can be obtained in a blast wave from a nanosecond breakdown \cite{marques2021over}. In the first approximation, this is a cylindrical wave propagating from the source - the waist of the nanosecond laser beam. The maximum increase in density inside the wave is determined by the factor $C=(\gamma+1)/(\gamma-1)$, here $\gamma$ is the adiabatic index, $\gamma=1.4$ for diatomic gases, and the density  increases by  $C\approx $ 6.  A region with a density an order of magnitude lower than the initial one appears between the source and the blast wave shock front. Thus, most of the medium from this region ends up at the shock front. Knowing the compression factor ($\approx6$), one can estimate the thickness of the blast wave's shell with radius $r$ as $dr\approx r/12$. Thus, the required density drop on the scale of 10-20 $\mu$m can be formed by the blast wave with  $r\approx$ 100-250 $\mu$m, which is reached within a few nanoseconds of its propagation. 

In this Letter, we for the first time utilized the cylindrical blast wave from nanosecond laser breakdown to control and probe LWFA of electrons by a 1 TW laser pulse in a subcritical plasma of a gas jet. It was demonstrated both experimentally and numerically that by shifting the shock front position along the acceleration channel one can stop the interaction of the electron bunch with the wake and extract an electron spectrum from the desired stage of acceleration. Furthermore, by properly choosing the shock front position a tunable quasi-monoenergetic electron beam with reduced divergence can be generated.

\begin{figure}
\includegraphics[width=\linewidth]{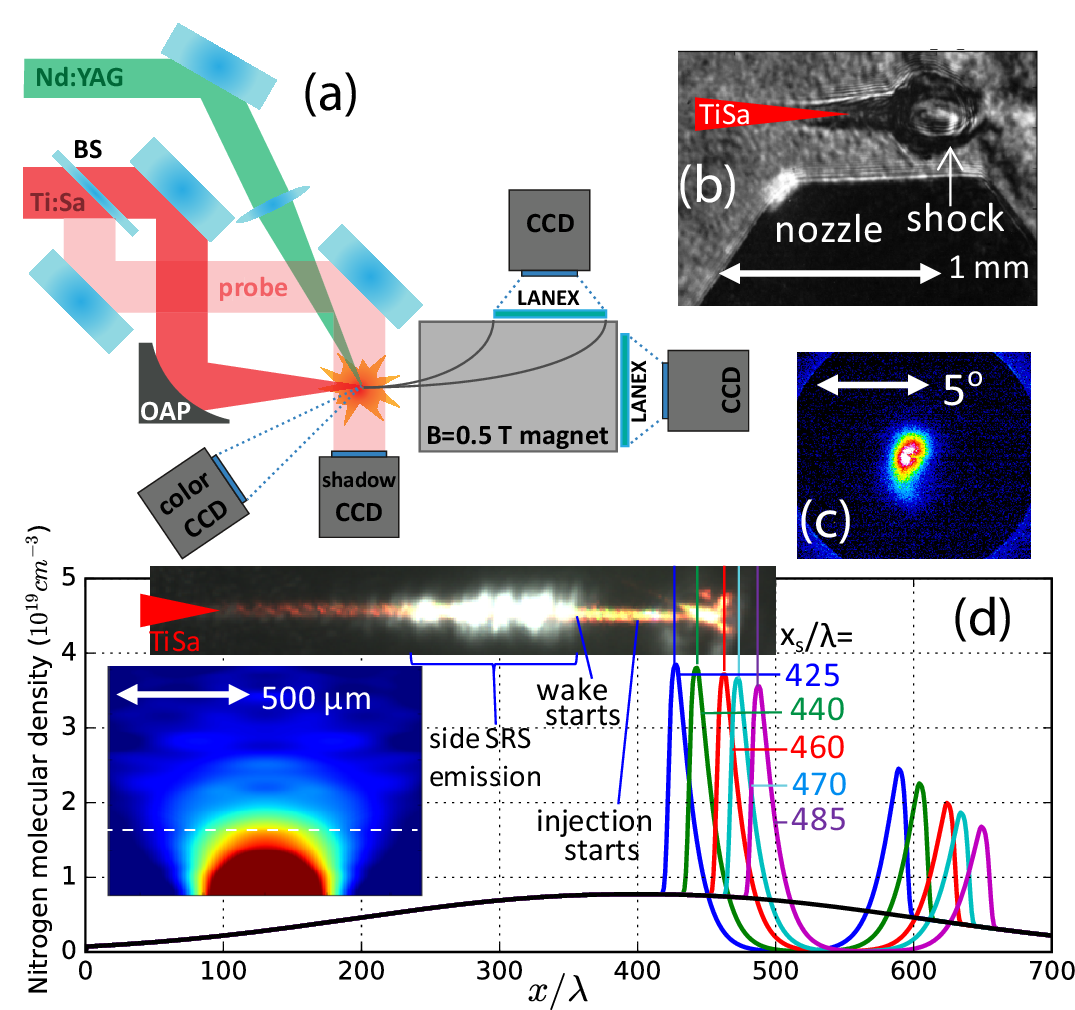}\caption{ 
\label{fig1}  (a) Experimental setup. The femtosecond Ti:Sa laser pulse (red beam) interacts with a gas jet modified by a blast wave from gas breakdown by an additional nanosecond Nd:YAG laser (green beam); (b) Shadowgraphy image of the jet with the blast wave and plasma channel from femtosecond beam; (c) The electron beam image; (d) 
Nitrogen molecular density in the blast wave for five different positions of its center (color lines, numbers -  left maxima of the shock front) with respect to the unpertubed density profile (black line) obtained at a height of 300 $\mu$m above the nozzle exit (white dashed line in the left inset showing density map from interferometric measurements). The upper inset to (d) shows image of radiation scattered from the plasma channel with the shock front at 460 $\mu$m (red curve in (d)).} 
\end{figure}

In our experiment (Fig.\ref{fig1}a) femtosecond laser pulse (Ti:Sapphire, $800$ nm, $50$ fs, $60$ mJ on target) was focused by an off-axis parabolic mirror ($f=15$ cm, estimated normalised vector potential $a_0=1.6$ in vacuum) into the point $x=100\lambda$ (x is the coordinate along the axis of beam propagation, the target center and the maximum density are located at $x=400\lambda$) of a continuous nitrogen gas jet at a height of 300 $\mu$m above the nozzle exit. The conical nozzle with an exit diameter of 400 $\mu$m at 2 bar backing pressure was used. The density profile obtained from interferometric measurements is shown in the inset to Fig.\ref{fig1}d, and the cross-section of the profile along the interaction line (white dotted line in the inset) is presented in Fig.\ref{fig1}d (black line). The cylindrical blast wave was created by the nanosecond laser pulse (Nd:YAG, 1064 nm, 10 ns, 200 mJ). This pulse was focused by a lens with an $f=15$ cm into the gas jet in the direction, perpendicular to the optical axis of the femtosecond beam, reaching a peak intensity of up to $10^{12}$ W/cm$^2$. Shadowgraphy image of this blast wave is shown in Fig.\ref{fig1}b. The position of the center of a blast wave along the optical axis of the femtosecond beam was changed by tilting the nanosecond laser beam with a motorized mirror located in front of the focusing lens.

Since interferometry cannot resolve sharp density spikes (by 6 times over 5 $\mu$m), numerical modeling of the interaction of the nanosecond pulse with nitrogen was made in the 3DLINE\cite{26Krukovskiy} package to deduce the molecular density distribution in the blast wave. The initial density distribution was taken from the interferometry (lower inset to Fig.\ref{fig1}d). Ionization kinetics was calculated using the Saha model, and radiation transport was calculated using the THERMOS\cite{27Vichev} code. The distributions in Fig.\ref{fig1}d (color lines) were obtained from numerical simulations 2 ns after the breakdown for the five different positions of the blast wave center ($x=505,520,540,550,565\lambda$, corresponding to the left shock front at $x=425,440,460,470,485\lambda$).

The femtosecond pulse ionizes the gas, undergoes self-focusing and channel formation occurs in the resulting plasma. The ionization region is visible in the shadow image (Fig.\ref{fig1}b). The plasma channel's emission is shown in the upper inset to Fig.\ref{fig1}d, detected with the color CCD camera ImagingSource DFK33GV024 equipped with the microscope lens (NA=0.2). The plasma channel scatters radiation at $\lambda=700-800$ nm (in addition, there is an ionization-induced blue shift) and appears red in this image. In the area $x=220-350\lambda$ an intense white glow appears. This radiation corresponds to Raman back and side scattering and it ceases when wakefield is excited. This diagnostic allows one to visualize the region of electron acceleration: the right boundary of the white glow near $x=350\lambda$ gives the point where the wakefield appears, and the position of the shock front determines the end point of acceleration.

\begin{figure}
\includegraphics[width=0.99\linewidth]{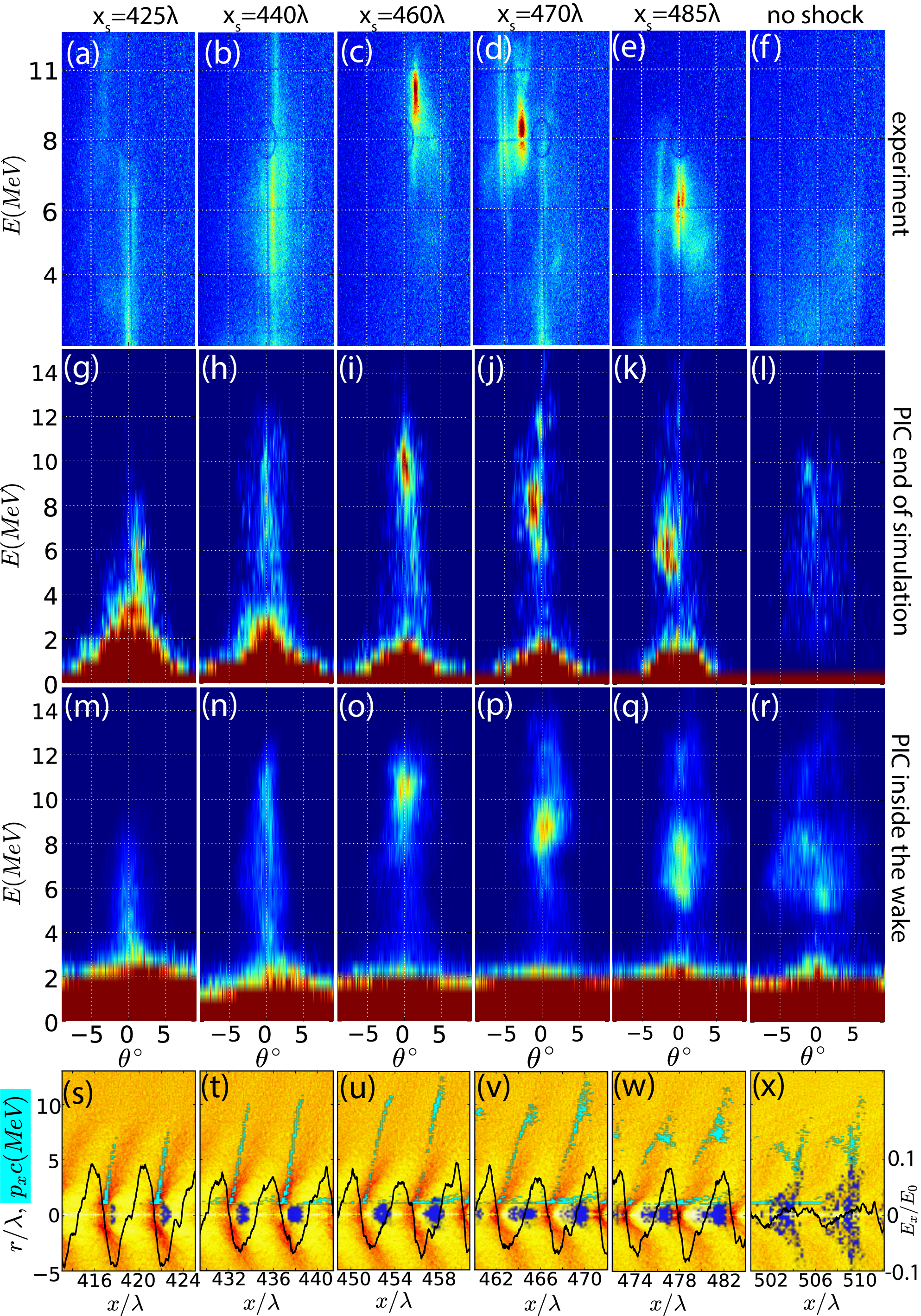}\caption{ 
\label{fig2} Electron spectra obtained in the experiment (a-f) and PIC simulations (g-l) for different shock front positions:  $x=425\lambda$ (a,g), $x=440\lambda$ (b,h), $x=460\lambda$ (c,i), $x=470\lambda$ (d,j), $x=485\lambda$ (e,k) and without shock (f,l). Spectra (g-l) show electrons that exited plasma (i.e. at time instant $t=50\lambda/c$, $t=0$ corresponds to the instant when the laser pulse enters plasma). Spectra m -- q correspond to g -- k but were plotted at the instants when the electron beam is in front of the shock ($t=30.6$, $31.2$, $32.4$, $32.9$, $33.4\lambda/c$, respectively). Spectrum \textit{r} was obtained at $t=35.6\lambda/c$ for the simulation without a blast wave. Images s -- x  show electron density $n_e$ (orange colormap), plasma longitudinal electric field $E_x$ (black line, normalized by $E_0=2\pi mc^2/\lambda e$), electrons distribution in phase space $x-p_x$ (cyan), accelerated electrons with energy $E > 6$ MeV (blue) obtained in PIC simulation at the same instants as m -- r without a blast wave. }
\end{figure}

Accelerated electrons were recorded by a LANEX screen with a filter cutting off low-energy electrons (2 mm Pb, electrons with $E>5$ MeV). The bunch charge (about $3$ pC) was assessed from the photoneutron yield using a 2 cm thick tungsten target introduced into the electron beam path (see details in \cite{gorlova2023neutron,Tsymbalov2019}). The minimum beam divergence was $1.5^{\circ}$ (Fig.\ref{fig1}c). Electron spectra were obtained at various positions of the shock front $x = 425, 440, 460, 470, 485\lambda$ using a 0.5 T magnetic spectrometer, see Fig.\ref{fig2}a--e. Fig.\ref{fig2}f shows the spectrum without the blast wave.

PIC simulations using SMILEI \cite{derouillat2018smilei} code in a quasi-3D (azimuthal symmetry) regime were also conducted for these six cases. A moving frame with a $0.98c$ speed, radius of $30\lambda$ (resolution $\lambda/8$), length of $60\lambda$ (resolution $\lambda/32$) and time step $\lambda/48c$ were used. Laser pulse had $\tau=50$ fs duration, field amplitude $a_0=1.6$ and was focused at $x=100\lambda$ with focal spot size $d=4\lambda$ as in the experiment. Molecular nitrogen density profiles obtained from the hydrodynamic simulations were used as a target (Fig.\ref{fig1}d). PIC simulations included ionization using the ADK model.

In the simulations, energy-angle distributions (hereafter referred to as 'spectrum') of the electron beam exiting the plasma were obtained for various positions of the shock front (Fig.\ref{fig2}g--k). These spectra correspond well with the spectra obtained for the same positions of the shock in the experiment (Fig.\ref{fig2}a--e). Figs.\ref{fig2}m-q display the spectra of electrons inside the wake just before passing through the shock front (the positions of the front are the same as in Fig.\ref{fig2}g-k). It is evident that the spectra of electrons inside the wake are quite close to those at the plasma exit if the acceleration is terminated by the shock front at a certain stage.  Therefore, the technique of tracking the evolution of the energy gain inside the wake was demonstrated by means of electron beam spectrum extraction at specific stage.

Figs.\ref{fig2}s--x illustrate the acceleration stages by showing electron density maps, longitudinal electric field, phase space of accelerated electrons, and electron bunches. The injection occurs on a smoothly decreasing density profile after the center of the nozzle. Electrons start to be trapped at $x=400\lambda$. At $t=30.6\lambda/c$, electrons in the first two buckets gain energy up to $7$ MeV (Fig.\ref{fig2}s). The experimental and numerical spectra obtained with the shock front at $x=425\lambda$  are shown in Figs.\ref{fig2}a,g, and the spectrum inside the wake at $t=30.6\lambda/c$ -- in Fig.\ref{fig2}m. These spectra are continuous. Figs.\ref{fig2}b,h show the spectra obtained with the shock at $x=440\lambda$. Here, electrons injected earliest gain an energy of $11$ MeV (Fig.\ref{fig2}n). The injection continues till the point $x=440\lambda$, while the spectrum remains continuous (Figs.\ref{fig2}n,t). Further to the point $x=440\lambda$, the injection weakens several times due to the decreasing amplitude of the plasma wave (decrease by 1.5 times can be seen in Figs.\ref{fig2}s--u). This occurs due to the energy loss of the laser pulse and the self-trapping of electrons. Injection termination creates a gap in the energy range $E<6$ MeV (see Figs.\ref{fig2}c,i,o,u).

Electrons are trapped from the crest of the plasma wave and quickly gain velocity close to the speed of light. Thus, an electron bunch of length $L\approx40\lambda(1-v_\phi/c)\approx\lambda\approx0.22\lambda_p$ ($v_\phi \approx 0.975c$ - wake phase velocity) is formed, determined by the injection duration ($\approx40\lambda$). The fastest electrons reach the middle of the bucket at $t=32.9\lambda/c$ and experience dephasing. Figs.\ref{fig2}p,q are plotted at $t=32.9,33.4\lambda/c$, when some of the electrons already experience decelerating field. 

Acceleration and deceleration would appear symmetric in 1D approximation provided the plasma wake did not change its parameters. Grouped around the maximum energies at time $t=32.4\lambda/c$, the beam would enter a decelerating field and stretch across the energy spectrum. The velocities immediately after the injection become close to $c$, and there is no significant electron re-grouping. Therefore, the electron bunch will look the same at the stages of acceleration and deceleration in a potential symmetrical with respect to the center of the bucket. However, the amplitude of the plasma wake during the acceleration length decreases by a factor of $\approx 1.5$. Thus the decelerating field has a lesser impact, resulting in the smaller electron bunch energy spread and formation of a quasi-monoenergetic spectrum. 

The optimal point to interrupt acceleration is evidently at $x=460\lambda$. The entire electron bunch is located in an accelerating focusing field near the center of the bucket. Since its length is $0.25\lambda_p$, and the field potential in the center of the bucket at a width of $0.25\lambda_p$ is almost constant, electrons at  $t=(32.4-32.7)\lambda/c$ gain maximum energies and minimum energy spread ($9-12$ MeV). Thus, the spectrum at $t=32.4\lambda/c$ (Fig.\ref{fig2}o) and, as a consequence, the spectrum obtained with the shock front at $x=460\lambda$ (Figs.\ref{fig2}c,i), turn out to be quasi-monochromatic with a peak at $8-11$ MeV. If we place the shock front at $x=470$ or $485\lambda$, a quasi-monochromatic beam with energies of $7-9$ MeV (Figs.\ref{fig2}d,j) or $5-7$ MeV (Figs.\ref{fig2}e,k) exits plasma. This beam has energies of $8-10$ and $6-8$ MeV prior to the shock front, respectively (Figs.\ref{fig2}p,v,q,w). 

\begin{figure}
\includegraphics[width=\linewidth]{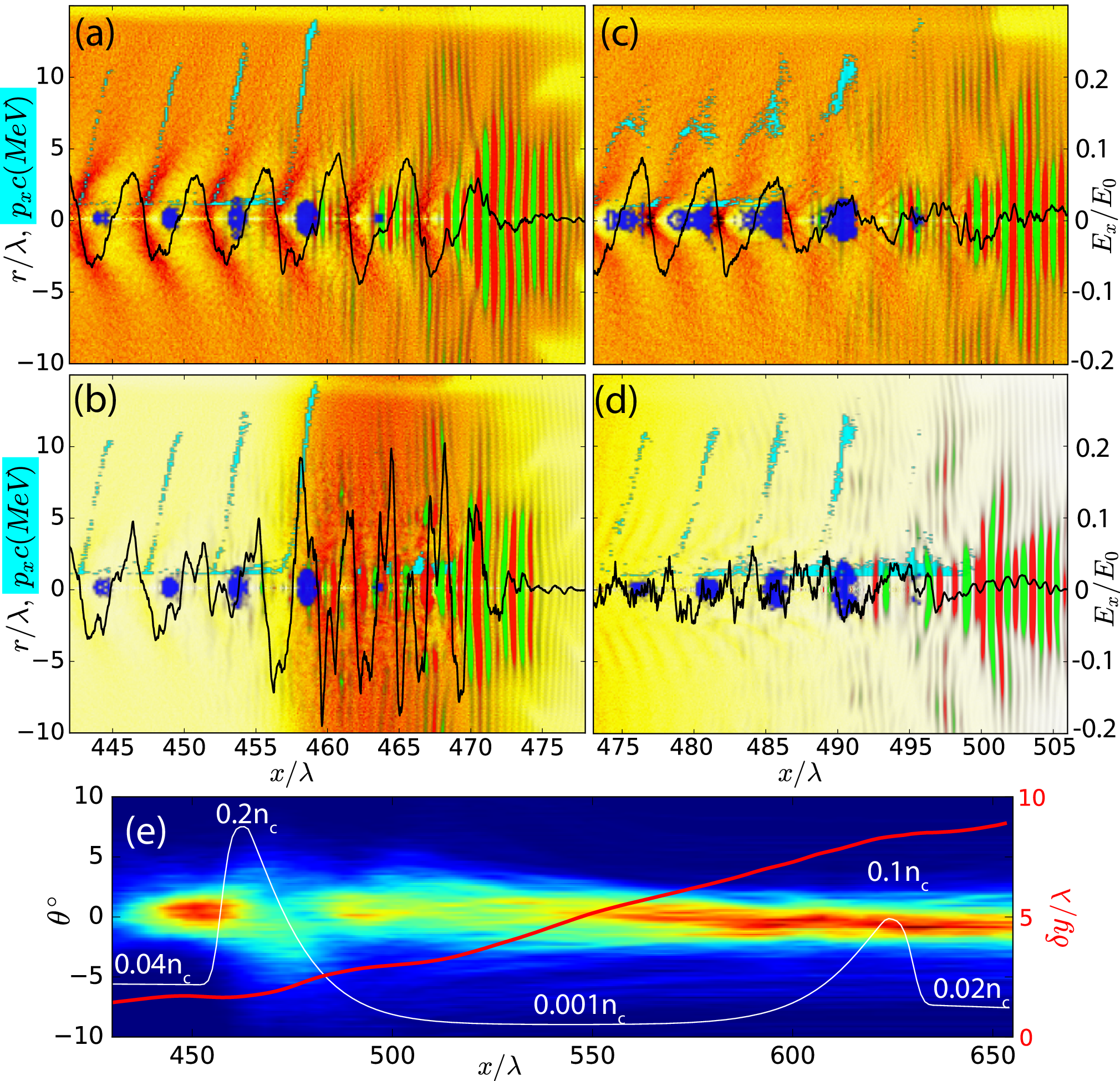}\caption{ \label{fig3} Electron density $n_e$ (orange colormap), plasma longitudinal electric field $E_x$ (black line, normalized by $E_0=2\pi mc^2/\lambda e$), electrons distribution in phase space $x-p_x$ (cyan), accelerated electrons $E > 6$ MeV (blue), laser pulse magnetic field (red and green for different polarities) obtained in the PIC simulations at $t=32.4\lambda/c$  (a,b), $t=34.4\lambda/c$ (c,d) without (a,c) and with (b, d) the shock; e - evolution of the angular spectrum of the beam ($E>6$ MeV) during propagation between the blast wave shock fronts (color map), diameter of the electron beam ($E>6$ MeV) (red line),  the profile of the electron density after ionization by the laser pulse (white line).}
\end{figure}

Comparing the spectra of electrons before the shock front (Figs.\ref{fig2}m--q) with that leaving the plasma (Figs.\ref{fig2}g--k), it is evident that energy reduction of $1-2$ MeV accompanied by the decrease in the beam divergence by a few times occurs. Let's analyze what happens to the beam when plasma is modified by the blast wave with its shock front located at $x=460\lambda$. The simulation data at $t=34.4\lambda/c$ without and with the shock is shown in Fig.\ref{fig3}. The electron density at the shock front increases by 5 times to $0.2n_c$, and the laser pulse excites plasma waves with $\lambda_p\approx2.5\lambda$ and an amplitude $\approx 3$ times higher than before the shock (Fig.\ref{fig3}b). 
Since these plasma waves oscillate rapidly and are localized inside the shock front ($10-20\lambda$) they change the energy of the electron beam insignificantly despite their higher amplitude. This can be observed by comparing phase space $x-p_x$ on Figs.\ref{fig3}b,d. By contrast, dephasing leads to lowering energies to $7$ MeV (Figs.\ref{fig3}a,c) without the shock at $t=34.4\lambda/c$. 

The decelerating field of the wake rapidly becomes defocusing, hence electrons acquire huge transverse momentum at the plasma exit (see Figs.\ref{fig2}f, l). The divergence of the electron beam in the accelerating phase is $\delta \phi\approx 3^{\circ}$ (Fig.\ref{fig2}m--o). If the beam interacts with the defocusing part of the wake its divergence increases up to $\delta \phi\approx 10^{\circ}$ (see Fig.\ref{fig2}r). 
This is not the case if the shock front is placed at smaller $x$ (Figs.\ref{fig2}i,j): beam with a divergence of $\delta \phi\approx 1.5^{\circ}$ is observed at the plasma exit. 

This is explained by the action of a plasma lens that is formed between the blast wave shock fronts. When passing through the shock front at $x=460\lambda$, the beam divergence increases to $10^{\circ}$ due to the action of the partially stochastic wake fields. The divergence almost restores after the shock at $x>480\lambda$ (Fig.\ref{fig3}e). Further, the laser pulse propagates in the rarefied plasma with $n\approx0.001-0.01n_{cr}$ ($x=500-600\lambda$). It expels electrons from the axis, thereby creating an ion channel with a predominantly radial electric field. This field does not change the energies of electrons, but slightly reduces electron beam divergence, while its size increases. The angular divergence is $\delta \phi_0 \approx 3^{\circ}$ and the beam transverse size is $\delta y_0 \approx 2.5\lambda$ just after the acceleration stage. Hence $\delta y\approx 6 \lambda$ at $x\approx 560\lambda$, i.e. before the right shock front (Fig. \ref{fig3}e).
Plasma density rises steadily to $0.002-0.02n_{cr}$ at $x>560\lambda$. Here the radial field becomes strong enough to focus the widened electron beam. As a result, the electron beam acquires the observed experimental divergence of $\delta \phi=1.5^{\circ}$ at $x=620\lambda$. Note, that in \cite{thaury2015demonstration,lehe2014laser} beam focusing was achieved by a plasma lens generated in an additional gas jet separated by free space from the main jet where the electrons were accelerated. We have created a similar configuration - acceleration - propagation with the increase in the beam transverse size  - focusing using the blast wave. 

In conclusion, we present a fully optical novel technique to control and diagnose the LWFA in a subcritical plasma. We applied a cylindrical blast wave with a density drop by an order of magnitude to control laser wakefield acceleration of electrons in an relatively dense plasma with $n\approx 0.05 n_c$. Blast wave shock front $\approx 10-20 \lambda$ thick was produced by the nanosecond laser pulse focused perpendicularly with respect to the femtosecond laser pulse accelerating electrons in the gas jet. The shock front can be placed at any point along the acceleration channel to stop the interaction of the accelerated electron bunch with the wake. Hence one can extract the electron bunch at the point of interest, avoiding dephasing and/or providing a diagnostic tool for the acceleration process.

In \cite{Cardenas_2020,faure2006controlled,swanson2017control} scanning along the acceleration length was achieved by moving the injection point. This method has essential drawbacks, as the wake evolves as it propagates, and changes in the injection point lead to variation in the accelerating field. Independent measurement can be achieved by adjusting the acceleration length through changes in the geometry of the gas cell, but this is possible on scales of millimeters and is suitable for GeV accelerators \cite{streeter2022characterization}. Recently Thomson scattering of a counter-propagating laser pulse on the accelerated electron beam was introduced as a diagnostic tool  \cite{Bohlen_2022}. This method requires precise spatial and temporal alignment of the two laser beams and allows for the measurement of the evolution of the maximum energy of the electron bunch only. It also cannot be used for electron energies below $10$ MeV due to the energy of the scattered X-ray quanta being too low. By contrast, our approach is suitable for arbitrary low-energy electrons (including the injection stage), does not imply any changes to the acceleration scheme, and is very easy to implement.

Our approach might also be useful for GeV-scale laser-plasma accelerators. The blast wave should create a low density space of sufficient length (up to a few cm) so that its right front reaches the end of the gas target and acceleration does not recommence. Increase in the diameter of the blast wave beyond 1 mm will result in a shock front that is too thick ($>100 \mu $m) and blurred \cite{marques2021over}. Therefore, for acceleration lengths exceeding a few mm it is advisable to use a cascade of blast waves or a longitudinally extended wave created by a flat laser beam along the longitudinal axis  
 \cite{seemann2023refractive}.

The cavity inside the blast wave not only allows the wake fields to be turned off but also focuses the electron beam. This can be used in multi-stage acceleration schemes, for example, to transition from LWFA to PWFA \cite{kurz2021demonstration}. The absence of the accelerating field inside the cavity of the blast wave can be analogous to drift sections in a RF accelerator and used to construct a multi-stage scheme similar to \cite{sadler2020overcoming} . 

The proposed technique was demonstrated experimentally and numerically in the case of self-modulated LWFA for the 50 fs, 1 TW laser pulse in the gas jet with peak electron density $n\approx 0.05n_c$ along the acceleration line. We showed how the quasi-monoenergetic electron beam can be formed despite dephasing. The first shock front destroys the wake thus terminating the acceleration, while a plasma lens formed between the two shock fronts reduces electron beam divergence by a factor of two. Consequently, a well-collimated ($\approx1.5^{\circ}$) tunable quasi-monoenergetic (4--11 MeV with $\approx$ 10\% energy spread) electron source was created.

This work was done with financial support from the RSCF (grant No. 22-79-10087). E.S. acknowledges the foundation for theoretical research BASIS for financial support. 

\bibliographystyle{apsrev4-2}
\bibliography{references}

\end{document}